\documentclass[
floatfix,
aps,
prl,
twocolumn,
superscriptaddress,
amssymb,
 groupaddress,
]{revtex4}

\usepackage{float}
\usepackage{amssymb}
\usepackage{gensymb}
\usepackage{latexsym}
\usepackage[pdftex]{graphicx}
\usepackage{amsmath}
\usepackage{graphicx}
\usepackage{dcolumn}
\usepackage{amsfonts}
\usepackage{bm}
\usepackage{epsfig}

\def\sin{Si$_{3}$N$_{4}\,$}
\def\alo{Al$_{2}$O$_{3}\,$}
\def\hfo{HfO$_{2}\,$}
\def\deg{^{\circ}}
\def\vg{V_{\mathrm{G}}}

\def\np{n_{+}}
\def\nn{n_{-}}
\def\nt{n_{\mathrm{T}}}
\def\lso{\ell_{\mathrm{SO}}}
\def\ar{\alpha_{\mathrm{r}}}

\def\oc{\omega_{\mathrm{c}}}
\def\tq{\tau_{\mathrm{q}}}
\def\mstar{\mathrm{m}^{\star}}

\def\nn{\nonumber}

\def\nt{\nu_{\mathrm{T}}}

\def\nt{n_{\mathrm{T}}}

\def\np{n_{+}}
\def\nn{n_{-}}
\def\nt{n_{\mathrm{T}}}
\def\lso{\ell_{\mathrm{SO}}}

\begin{document}
\title{High mobility InAs 2DEGs on GaSb substrates: a platform for mesoscopic quantum transport}

\author{C. Thomas}
\thanks{These two authors contributed equally.}
\affiliation{Station Q Purdue and Department of Physics and Astronomy, Purdue University, West Lafayette, Indiana 47907, USA}
\affiliation{Birck Nanotechnology Center, Purdue University, West Lafayette, IN, 47907 USA}

\author{A.\,T. Hatke}
\thanks{These two authors contributed equally.}
\affiliation{Station Q Purdue and Department of Physics and Astronomy, Purdue University, West Lafayette, Indiana 47907, USA}

\author{A. Tuaz}
\affiliation{Station Q Purdue and Department of Physics and Astronomy, Purdue University, West Lafayette, Indiana 47907, USA}
\affiliation{Birck Nanotechnology Center, Purdue University, West Lafayette, IN, 47907 USA}

\author{R. Kallaher}
\affiliation{Birck Nanotechnology Center, Purdue University, West Lafayette, IN, 47907 USA}
\affiliation{Microsoft Quantum at Station Q Purdue, Purdue University, West Lafayette, Indiana 47907, USA}

\author{T. Wu}
\affiliation{Station Q Purdue and Department of Physics and Astronomy, Purdue University, West Lafayette, Indiana 47907, USA}

\author{T. Wang}
\affiliation{Station Q Purdue and Department of Physics and Astronomy, Purdue University, West Lafayette, Indiana 47907, USA}
\affiliation{Birck Nanotechnology Center, Purdue University, West Lafayette, IN, 47907 USA}

\author{R. E. Diaz}
\affiliation{Birck Nanotechnology Center, Purdue University, West Lafayette, IN, 47907 USA}

\author{G. C. Gardner}
\affiliation{Birck Nanotechnology Center, Purdue University, West Lafayette, IN, 47907 USA}
\affiliation{Microsoft Quantum at Station Q Purdue, Purdue University, West Lafayette, Indiana 47907, USA}

\author{M. A. Capano}
\affiliation{Birck Nanotechnology Center, Purdue University, West Lafayette, IN, 47907 USA}
\affiliation{School of Electrical and Computer Engineering, Purdue University, West Lafayette, Indiana 47907, USA}

\author{M. J. Manfra}
\affiliation{Station Q Purdue and Department of Physics and Astronomy, Purdue University, West Lafayette, Indiana 47907, USA}
\affiliation{Birck Nanotechnology Center, Purdue University, West Lafayette, IN, 47907 USA}
\affiliation{School of Electrical and Computer Engineering, Purdue University, West Lafayette, Indiana 47907, USA}
\affiliation{School of Materials Engineering, Purdue University, West Lafayette, IN, 47907 USA}

\received{\today}

\begin{abstract}
High mobility, strong spin-orbit coupling, and large Land{\'e} g-factor make the two-dimensional electron gas (2DEG) in InAs quantum wells grown on nearly-lattice-matched GaSb substrates an attractive platform for mesoscopic quantum transport experiments. 
Successful operation of mesoscopic devices relies on three key properties: electrical isolation from the substrate; ability to fully deplete the 2DEG and control residual sidewall conduction with lithographic gates; and high mobility to ensure ballistic transport over mesoscopic length scales.  
Simultaneous demonstration of these properties has remained elusive for InAs 2DEGs on GaSb substrates.  
Here we report on heterostructure design, molecular beam epitaxy growth, and device fabrication that result in high carrier mobility and full 2DEG depletion with minimal residual edge conduction.  
Our results provide a pathway to fully-controlled 2DEG-based InAs mesoscopic devices.
\end{abstract} 

\maketitle

\section{Introduction}

With low effective mass, strong spin-orbit coupling and high Land{\'e} g-factor \citep{Lutchyn:2018}, the two-dimensional electron gas (2DEG) in InAs quantum wells is an interesting platform for mesoscopic physics experiments, but only a few realizations of mesoscopic devices have been reported with trench-etched quantum point contacts, e.g. \cite{Debray:2009,shabani:2014}.
More recently, lithographically-defined nanowires comprised of the s-wave superconductor aluminum proximity-coupled to InAs 2DEG grown on semi-insulating InP have experimentally revealed evidence of Majorana zero modes \citep{nichele:2017,suominen:2017}.
Motivated by the possibility to explore mesoscopic phenomena and novel topological states in a lower disorder environment, significant effort is now dedicated to improvement of InAs 2DEG electronic properties.
For example, the growth of composite quantum wells of InGaAs/InAs \citep{wallart:2005} enabled improvement of the carrier mobility to $\mu = 1.0 \times 10^{6}\,$cm$^{2}/$Vs for InAs quantum wells buried $120\,$nm below the top surface and grown on semi-insulating (100) InP substrates \citep{hatke:2017}. 
However, growth of InAs on InP substrates must accommodate the nearly $3.5\%$ lattice mismatch to InAs through the use of thick graded buffer layers that introduce dislocations and modify surface morphology.
%AH1: slight rewording
By growing InAs quantum wells on quasi-lattice-matched (100) GaSb substrates, 2DEG mobility has reached $\mu \sim 2.4 \times 10^{6}\,$cm$^{2}/$Vs at density $n \sim1\times10^{12}$cm$^{-2}$ \citep{tschirky:2017}, for a quantum well buried $25\,$nm from the top surface, through reduced defect generation and use of AlGaSb barriers that increase electron wavefunction confinement.  
Many body effects associated with the fractional quantum Hall effect \citep{ma:2017} have been recently reported on such structures, indicative of improved quality.

However, mesoscopic experiments using high quality InAs 2DEGs on GaSb substrates still present challenges. 
Specific limitations include: 1) parasitic conduction through the substrate; due to the small bandgap of GaSb, substrates are not perfectly insulating which may limit operation of high impedance quantum devices in the tunnelling regime; 2) the inability to fully deplete the 2DEG and residual sidewall conduction with surface gates. 
High mobility 2DEGs utilizing a GaSb buffer layer have been successfully grown \citep{tschirky:2017}. 
However simple GaSb buffers are not expected to result in sufficient isolation from the GaSb substrate. 
Growth of lattice matched buffers composed of higher bandgap materials such as AlSb$_{y}$As$_{1-y}$ \citep{sho:2016,shojaei:2018} or Al$_{x}$Ga$_{1-x}$Sb$_{y}$As$_{1-y}$ \citep{shojaei:2016} appears promising to address these issues but no data have been published detailing device isolation or full 2DEG depletion.
Additionally, it has been reported that high mobility InAs 2DEGs suffer from excess conductance attributed to side wall conduction, most probably associated with charge accumulation at the exposed InAs edge. 
Mueller \textit{et al.} \citep{mueller:2017} reported edge resistivity of 1.3-2.5 k$\Omega$/$\mu$m. 
Edge resistivity was subsequently increased to $72.8$ k$\Omega/\mu$m  by Mittag and coworkers by incorporating an Al$_2$O$_3$ gate dielectric deposited by atomic layer deposition (ALD) on an heterostructure designed to host a low density ($n \sim 5 \times 10^{11}$cm$^{-2}$) InAs 2DEG \citep{mittag:2017}.

Here we report on the design, fabrication, and operation of devices based on InAs 2DEGs on (100) GaSb substrates that address the aforementioned limitations. 
To better understand the impact of epilayer design on device performance, a series of heterostructures was grown by molecular beam epitaxy with various buffer and cap layer materials. Buffer layers must electrically isolate the 2DEG from the substrate and cap layers that terminate the semiconductor stack must promote strong electrostatic gate control.  
Our principal results can be summarized as follows. With its large bandgap, a buffer layer of the quaternary compound Al$_{0.8}$Ga$_{0.2}$Sb$_{0.93}$As$_{0.07}$ is found to effectively isolate the 2DEG from the GaSb substrate and yields mesa device isolation larger than $1\,$G$\Omega$.  
Moreover, the precise lattice-match of Al$_{0.8}$Ga$_{0.2}$Sb$_{0.93}$As$_{0.07}$ with GaSb enables the growth of InAs quantum wells on a dislocation free buffer resulting in electron mobility      
 $\mu=2.1\times10^6$ cm$^2/$Vs at density $n=6.7\times10^{11}$ cm$^{-2}$.
A thin InAs cap layer has enabled complete depletion of the 2DEG with minimal residual sidewall conduction and has proved to be favorable in this context over the commonly used GaSb \citep{tschirky:2017, shojaei:2016}. 
Interestingly, the choice of an InAs or GaSb capping layer dramatically impacts scattering in the 2DEG. The GaSb cap layer appears to screen the 2DEG from surface defect scattering, enhancing mobility, but simultaneously prevents full 2DEG depletion.  The InAs cap layer does not screen surface charge as efficiently, but does facilitate full 2DEG depletion.

\section{Molecular beam epitaxy growth and device fabrication}
%%%%%%%%%%%%%%%%%%%%%%%%%%%%%%%%%%%%%%%%%%%%%%%%%%%%%%
\begin{table*}[t]
  \scalebox{0.95}{
  \begin{tabular}{ | c | c | c | c | c | c | }
    \hline
    \,\, Sample \,\, & \,\,Buffer Layer\,\, &\,Cap Layer (Thickness)\,& \,\,\sin\,\, & \,\,\alo\,\, & \,\,\hfo\,\, \\ \hline
    \,\,A\,\, & \,GaSb\, & GaSb ($2\,$nm) & \,10-100 k$\Omega$\, & \,\, ----- \,\, & \,\, ----- \,\, \\ \hline
    \,\,B\,\, & \,GaSb/AlSb SL\, & \,GaSb ($2\,$nm)\, & 1 M$\Omega$ & \,\, ----- \,\, & \,\, ----- \,\, \\ \hline
    \,\,C\,\, & \,Al$_{0.35}$Ga$_{0.65}$Sb\, & \,GaSb ($2\,$nm)\, & \,\, ----- \,\, & 1 M$\Omega$ & \,\, ----- \,\,  \\ \hline
    \,\,D\,\, & \,AlSb$_{0.91}$As$_{0.09}$\, &  \,GaSb ($2\,$nm)\, & \,\, ----- \,\, & 1 M$\Omega$ & \,\, ----- \,\,  \\ \hline
    \,\,E\,\, & \,Al$_{0.8}$Ga$_{0.2}$Sb$_{0.93}$As$_{0.07}$\, & \,Al$_{0.8}$Ga$_{0.2}$Sb$_{0.93}$As$_{0.07}$ ($3\,$nm)\, & $>$G$\Omega$ & $>$G$\Omega$ &  $>$G$\Omega$ \\ \hline
    \,\,F\,\, & \,Al$_{0.8}$Ga$_{0.2}$Sb$_{0.93}$As$_{0.07}$\, & \,GaSb ($2\,$nm)\, & \,\, ----- \,\, & \,\, ----- \,\, &  $>$G$\Omega$ \\ \hline
    \,\,G\,\, & \,Al$_{0.8}$Ga$_{0.2}$Sb$_{0.93}$As$_{0.07}$\, & \,InAs ($2\,$nm)\, & \,\, ----- \,\, & \,\, ----- \,\, &  $>$G$\Omega$ \\
    \hline
  \end{tabular}}
  \caption{Mesa-to-mesa isolation resistance between etched Hall bars.
Buffer layer column is the terminated layer composition to which the Hall bars are etched and cap layer column is the material that terminates the surface. 
Finally, the 4$^{\mathrm{th}}$ - 6$^{\mathrm{th}}$ columns are the order of magnitude resistances between etched mesas for different applied dielectrics.}
\label{m2m_table}
\end{table*}
%%%%%%%%%%%%%%%%%%%%%%%%%%%%%%%%%%%%%%%%%%%%%%%%%%%%%%
InAs-based 2DEGs (see Fig.\,\ref{xrd_tem}(a) for a typical layer stack) were grown on undoped GaSb (100) substrates by molecular beam epitaxy in a Veeco Gen 930 using ultra-high purity techniques and methods as described in Ref.\,\onlinecite{gardner:2016}. 
Substrate temperature was measured by blackbody radiation emission. 
To improve epilayer morphology, the native oxide of the substrate was removed at  $400 \deg\,$C using atomic hydrogen without Sb flux. 
This process is found to improve surface morphology significantly by limiting the formation of surface defects generated during thermal oxide desorption \cite{Weiss:2007}. 
The substrate temperature was then increased to $500 \deg\,$C, under Sb flux and growth was initiated with a $25\,$nm smoothing layer of GaSb followed by a thick buffer layer of $800\,$nm.  
Five samples (A - E) were grown to investigate buffer layers of varying complexity ranging from a pure GaSb to Al$_{0.8}$Ga$_{0.2}$Sb$_{0.93}$As$_{0.07}$ lattice matched to GaSb, as summarized in Table\,\ref{m2m_table}. 
Precise control of substrate temperature is of principal importance for the ternary AlSb$_{y}$As$_{1-y}$ (Sample D) and the quaternary Al$_x$Ga$_{1-x}$Sb$_{y}$As$_{1-y}$ (Sample E) layers since it directly controls the incorporation ratio of Sb and As and thus the lattice parameter. Growth of Al(Ga)SbAs compounds (Samples D and E) following the hydrogen-assisted oxide removal process enables smooth buffer layers to be grown free of pyramidal defects typically observed on the surface of GaSb grown on GaSb \citep{murray:2013}.

The active region of the heterostructure is composed of $24\,$nm of InAs flanked by Al$_{x}$Ga$_{1-x}$Sb barriers with $x= 0.8$. 
The top barrier thickness is $20\,$nm unless otherwise specified. 
Finally, the growth was terminated by deposition of a cap layer, whose primary role is to avoid oxidation of the underlying Sb-containing layers and prevent formation of parasitic conduction channels. 
As we shall demonstrate, the cap layer plays a crucial role for electrostatic control of the 2DEG. 
GaSb, AlGaSbAs, and InAs cap layers were explored in three different samples (E - G).
For all samples in this study, transitions between InAs and Sb-containing compounds made use of a shutter sequence pioneered in Ref.\,\onlinecite{tuttle:1990}. 
The growth of InAs was performed with an effective As-to-In ratio slightly larger than 1 to prevent the formation of void defects associated with As etching of the Sb-based layer underneath. 
Under these conditions, flat surface morphology was obtained with a roughness on the order of a monolayer.
Throughout the entire growth the Sb flux was kept constant with a beam equivalent pressure of $4.0 \times 10^{-7}\,$torr.

%\subsection{Processing}

Devices were processed with standard wet etching techniques to define straight and L-shaped Hall bars of width $w=50$ and $150\,\mu$m to measure transport along different crystallographic directions.
To avoid complications arising from oxidation of etched surfaces, we first deposit Ohmic Ti/Ni/Au $(20/200/100\,$nm$)$ contacts.
Our samples are then etched with a H${_2}$O\,:\,citric monohydrate\,:\,H$_{3}$PO$_{4}$\,:\,H$_{2}$O$_{2}$\,(220 mL: 55 g: 3 mL: 5 mL) solution to define the Hall bar mesas, where the etch is terminated in the buffer layer. 
After deposition of a dielectric of \alo or \hfo using thermal ALD at $150 \deg\,$C and $90 \deg\,$C, respectively, or \sin using plasma enhanced chemical vapor deposition at $250 \deg\,$C, a Ti/Au $(20/150\,$nm$)$ gate is deposited.
Processing is completed with deposition of a Ti/Ni/Au $(20/200/100\,$nm$)$ bond pad after a BOE etch to open a window in the dielectric over the Ohmic contacts.

\section{Results}
\subsection{Device isolation}
Samples were measured in a $^{3}$He system at a base temperature of $T=300\,$mK using standard low frequency lock-in and dc techniques.  
We first discuss the impact of heterostructure design on device electrical isolation.  
Mesoscopic devices must be well isolated from the substrate in order to operate in a high impedance configuration.  
We test the degree of isolation by applying dc voltage bias between Ohmic contacts on neighboring mesas and monitoring the current that flows between them.

The resistance between different etched Hall bars was measured by floating all contacts on two adjacent Hall bars except one test contact on each mesa and applying a voltage between the two mesas.
This configuration forces any current flow between the mesas to, presumably, occur either at the interface of the etched semiconductor layer and the dielectric or down through the buffer layers into the weakly conducting substrate and back up through the buffer again to the other mesa.
We investigated I-V curves in the range $-0.5\,$V$<V<0.5\,$V, where $V$ is an applied voltage between Ohmic contacts.
The mesa-to-mesa resistances for different buffer layer and dielectric combinations are summarized in Table \ref{m2m_table} for Samples A\,-\,G,. 
In each case, the mesa etch is terminated in the buffer layer.
For resistances labelled as $>$G$\Omega$ no current is observed above the noise floor of our measurement circuit.  As the current noise floor is approximately $0.5\,$pA, $1\,$G$\Omega$ is a lower bound on the actual resistance.

%%%%%%%%%%%%%%%%%%%%%%%%%%%%%%%%%%%%%%%%%%%%%%%%%
\begin{figure*}[t]
\includegraphics[scale=0.37]{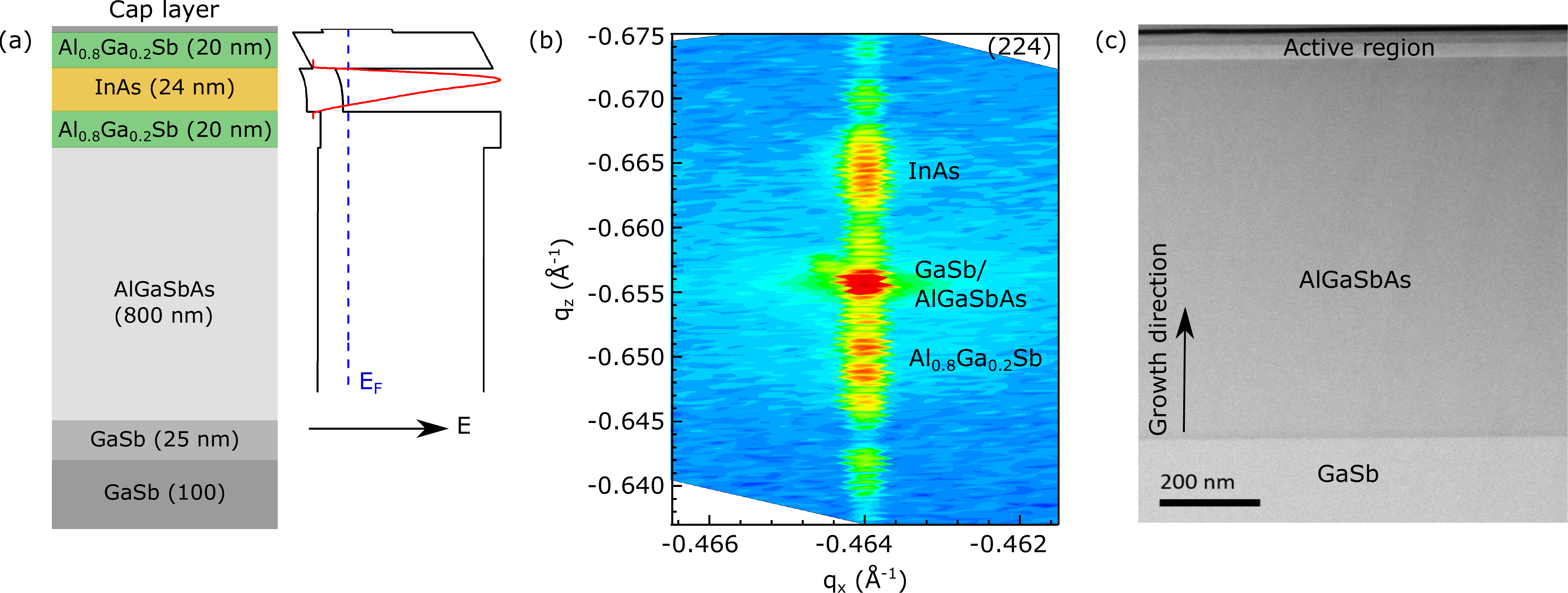}
\caption{(Color online)
(a) Layer stack of Samples E to G with the associated bandstructure of Sample F generated using Nextnano \cite{nextnano}. 
The black bands represent the conduction and valence bands at the $\Gamma$ point. 
The dashed blue line is the Fermi level denoted $E_F$ and the red curve corresponds to the two dimensional electron system wavefunction. 
(b) High resolution X-ray diffraction reciprocal space map in the asymmetric  (224) direction for Sample E. 
(c) Scanning transmission electron micrograph in high angle annular dark field mode with $[110]$ zone axis focusing on the AlGaSbAs buffer layer for Sample E.
}
\label{xrd_tem}
\end{figure*}
%%%%%%%%%%%%%%%%%%%%%%%%%%%%%%%%%%%%%%%%%%%%%%%%%

We first consider Sample A, where GaSb has been used as the buffer layer, similarly to Ref.\,\onlinecite{tschirky:2017}. 
A low resistance between mesas on the order of $10-100\,$k$\Omega$ is measured with \sin as a dielectric. 
For Sample B, the GaSb buffer is terminated with a superlattice (SL) made of $1\,$nm-thick GaSb and $1\,$nm-thick AlSb layers, repeated 100 times.  
With the same \sin dielectric, the incorporation of the SL results in an increase of the mesa-to-mesa resistance to the order of a few M$\Omega$s. 
While this constitutes a significant improvement over the bare GaSb buffer, the resistance is still not sufficient to isolate nanostructures as efficiently as a semi-insulating substrate. 
Sample C is comprised of an Al$_{0.35}$Ga$_{0.65}$Sb buffer that has a larger bandgap ($1.2\,$eV \citep{Shim:2015}) but is not well lattice-matched (mismatch of $0.23\%$) in comparison with the GaSb buffer of Sample A.  
Sample D has a buffer made of the ternary alloy AlSb$_{0.91}$As$_{0.09}$ that is characterized by a bandgap twice as large ($2.4\,$eV \citep{Shim:2015}) and better lattice matching to GaSb (mismatch $\lesssim -0.1\%$) than Sample C. 
For these samples, \alo is used as dielectric to reduce the thermal budget during processing.
Despite the nominally better intrinsic properties of these two buffer materials, we still measured a resistance of a few M$\Omega$s.
Sufficient isolation, $>$G$\Omega$, was only obtained by increasing the complexity of the buffer layer to the quaternary Al$_{0.8}$Ga$_{0.2}$Sb$_{0.93}$As$_{0.07}$ alloy, which is used for Samples E\,-\,G. 
Note that Al$_{0.8}$Ga$_{0.2}$Sb$_{0.93}$As$_{0.07}$ is fully lattice-matched to the underlying GaSb substrate. 
We additionally infer that the addition of gallium and arsenic to the buffer limits the oxidation of the exposed buffer layer surface after the mesa etch.  
For Sample E, three chips were prepared with the three different dielectric materials: \sin, \alo and \hfo and all gave device isolation  $>$G$\Omega$. 
In the remainder of this manuscript, the measurements presented for Sample E are on devices having \hfo as dielectric. 
These results highlight the primary importance of the epitaxial buffer layer material over the particular dielectric used for improvement of device isolation. 

\subsection{Structural characterization}
With the validation of the electronic properties of the Al$_{0.8}$Ga$_{0.2}$Sb$_{0.93}$As$_{0.07}$ buffer, we comment on its structural properties. 
To determine the strain state and test for possible relaxation in these heterostructures, high resolution x-ray reciprocal space mapping (RSM) has been performed using a X'pert PANalytical diffractometer. 
Performing RSM along an asymmetric reflection provides insight into the in- and out-of-plane lattice parameters and thus enables extraction of the strain profile of the different layers of our heterostructures. 
Figure\,\ref{xrd_tem}\,(b) displays an RSM centered on the GaSb asymmetric $(224)$ diffraction peak using a glancing exit configuration. 
The mapping has been collected over a $2\theta-\omega$ range of 4$\degree$ with $0.02\degree$ steps and an $\omega$ range of $1.6\degree$ with steps of $0.02\degree$. 
Note that with the use of a copper X-ray tube, an X-ray mirror and a channel-cut Ge $(2 2 0)$ crystal, photons were provided at an energy of $8.048\,$keV and a divergence of $0.008\degree$. 
The  acceptance angle of the detector was $0.2\degree$.   
The Al$_{0.8}$Ga$_{0.2}$Sb$_{0.93}$As$_{0.07}$ and GaSb peaks cannot be resolved which indicates good lattice matching of the buffer to the substrate. 
All peaks share a common in-plane scattering vector ($q_x= -0.4640$ {\AA}$^{-1}$) that confirms the constituent layers possess the same in-plane lattice parameter. The vertical alignment of peaks in the RSM demonstrates all layers are coherently strained and there is no evidence of relaxation.
The crystalline quality of this heterostructure is confirmed by scanning transmission electron microscopy (STEM). STEM was performed using a Thermo Scientific Talos 200X TEM and images were acquired with a high angle annular dark field along the $[110]$ zone axis. The STEM image of Fig.\,\ref{xrd_tem}(c) highlights a uniform AlGaSbAs buffer layer with no evidence of threading dislocations. 
%The crystalline quality of this heterostructure is confirmed by scanning %transmission electron microscopy (STEM), operated in high angle annular dark %field mode along the $[110]$ zone axis. 
%The scanning transmission electron micrograph of Fig.\,\ref{xrd_tem}(c) %highlights a uniform AlGaSbAs layer with no evidence of threading dislocations. 

%%%%%%%%%%%%%%%%%%%%%%%%%%%%%%%%%%%%%%%%%%%%%%%%%

%Saturation of $n$ data comes from Sample E (M12-13-17.1).

%%%%%%%%%%%%%%%%%%%%%%%%%%%%%%%%%%%%%%%%%%%%%%%%%
\begin{figure}[t]
\vspace{-0.1 in}
\includegraphics[width=0.5\textwidth]{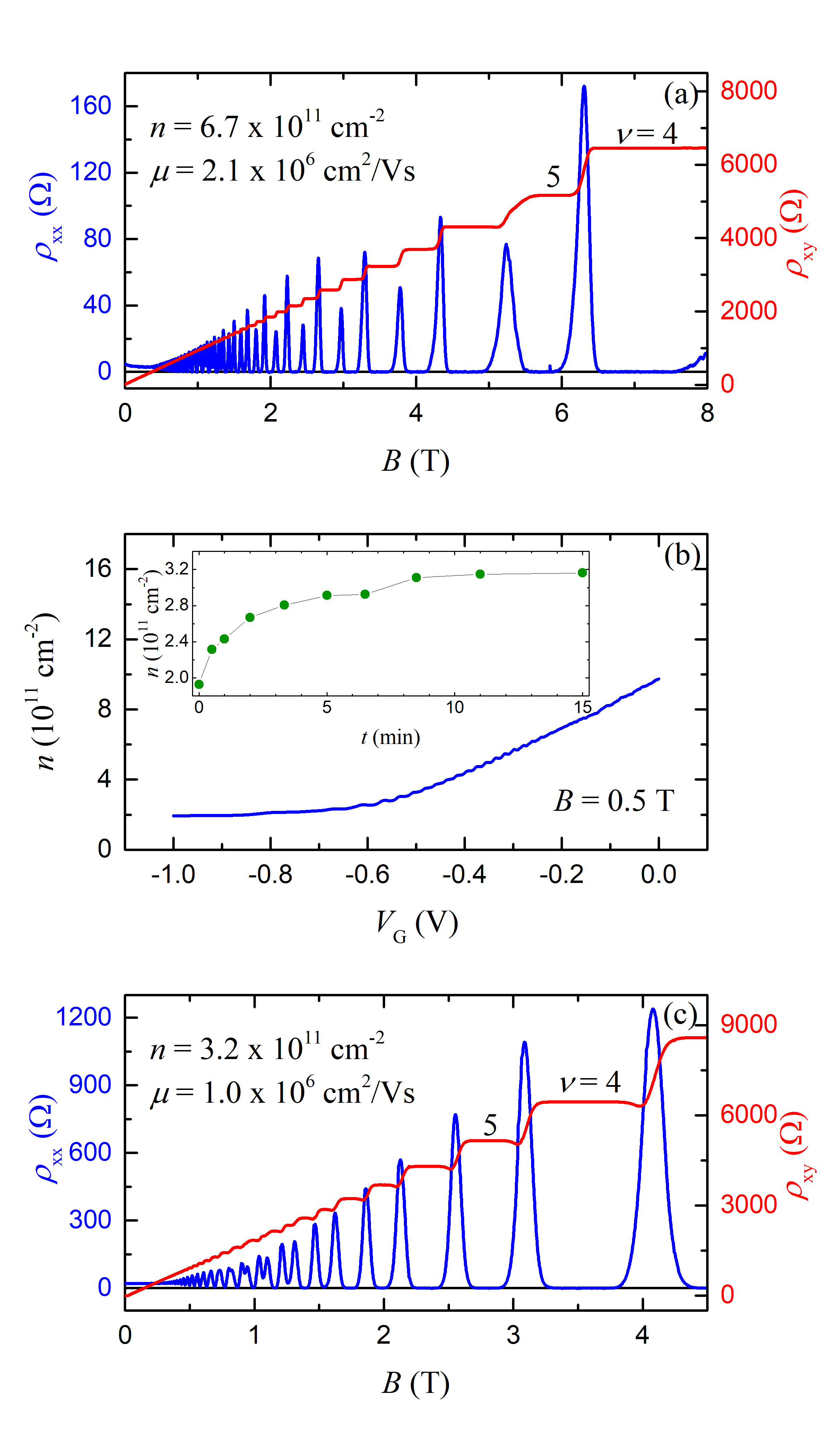}
\vspace{-0.25 in}
\caption{(Color online)
(a) Magnetoresistance of Sample E at $n = 6.7 \times 10^{11}\,$cm$^{-2}$ near peak mobility of $\mu = 2.1 \times 10^6\,$cm$^{-2}$/Vs.
(b) Density $n$ vs gate voltage $\vg$ obtained at $B=0.5\,$T from Sample E by sweeping $\vg =0\,$V $\rightarrow$ $-1\,$V at $2\,$mV/s.
Inset: $n$ vs time $t$ at $\vg=-1.1\,$V.
(c) Magnetoresistance of Sample E at $\vg=-1.1\,$V after 15 minute equilibration.
} 
\label{algaassb_cap}
\end{figure}
%%%%%%%%%%%%%%%%%%%%%%%%%%%%%%%%%%%%%%%%%%%%%%%%%

\subsection{Carrier depletion}
The second identified challenge for InAs/GaSb heterostructures is the inability to fully deplete the carriers in fabricated devices. 
Given that we employ high bandgap materials for both the barriers surrounding the quantum well and the buffer (see the bandstructure of Fig. \ref{xrd_tem}(a)), we focus our attention on the capping layer as the possible location of bands crossing the Fermi level that may limit full depletion of the 2DEG.
Selection of GaSb as the cap layer (Sample F) is natural because it is lattice matched to the structure and Ga compounds are known to inhibit oxide formation better than Al-rich compounds, such as those used as our top barrier. This choice is widely reported with various GaSb capping layer thicknesses \citep{tschirky:2017,sho:2016}. Full carrier depletion, however, was not observed in our experiments. This observation is corroborated by Nextnano simulations \citep{nextnano} that suggest an absence of a clear depletion region without hole accumulation.

In Fig.\,\ref{algaassb_cap}\,(a) we present magnetotransport data for Sample E that employs a higher bandgap $3\,$nm-thick Al$_{0.8}$Ga$_{0.2}$Sb$_{0.93}$As$_{0.07}$ cap layer.
Note that, at most, a $5\%$ anisotropy between the $[011]$ and $[01\bar{1}]$ crystallographic directions has been detected for all samples so only the $[01\bar{1}]$ direction will be considered for the remainder of the manuscript.
Here we observe well defined quantum plateaus of the Hall resistance $\rho_{xy}$ with corresponding vanishing minima of the longitudinal resistance $\rho_{xx}$ indicative of high quality transport without significant parallel conduction.

Despite the excellent transport and very high 2DEG mobility, this device does not fully deplete as demonstrated in Fig.\,\ref{algaassb_cap}\,(b).
Starting from zero gate voltage, a linear relation between the carrier density $n$ and the gate voltage $\vg$ is observed for $-0.5 <\vg\leq 0\,$V and would extrapolate to full depletion of the device at $\vg\sim-0.7\,$V. 
However, for $\vg \leq -0.5$V, the density is no longer linearly varying with $\vg$. 
A saturation of $n$ is instead observed for $\vg \leq -0.7\,$V with $n \sim 2.0 \times 10^{11}\,$cm$^{-2}$. 
The $\vg$ sweep is truncated at $\vg=-1\,$V. 

%%%%%%%%%%%%%%%%%%%%%%%%%
\begin{figure}[t]
\vspace{-0.1 in}
\includegraphics[width=0.48\textwidth]{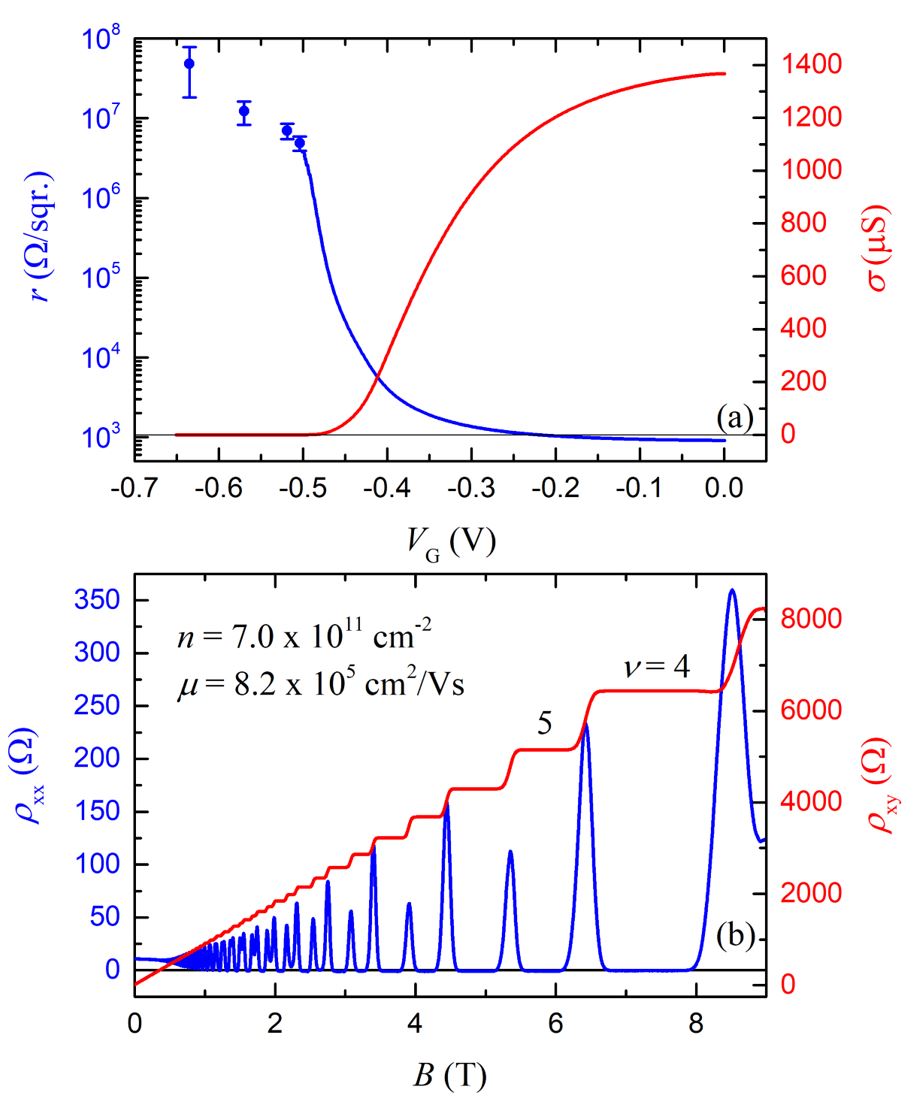}
\vspace{-0.25 in}
\caption{(Color online)
(a) Gate voltage dependence of the resistance per square (left axis, semi-log scale) and conductivity (right axis) through a straight Hall bar fabricated from Sample G.
(b) Magnetotransport near mobility peak $\mu = 8.2 \times 10^5\,$cm$^{2}$/Vs for  $n = 7.0 \times 10^{11}\,$cm$^{-2}$ for Sample G.
}
\label{inas_cap}
\end{figure}

%%%%%%%%%%%%%%%%%%%%%%%%%

Within the saturation regime, $n$ is not stable and increases with time $t$, as indicated in the inset of Fig.\,\ref{algaassb_cap}\,(b). 
After $t\sim 15\,$min, the density equilibrates at $n \sim 3.2 \times 10^{11}\,$cm$^{-2}$, which represents an increase of  $ \sim 1.2 \times 10^{11}\,$cm$^{-2}$ at fixed gate voltage.
Once $n$ is stabilized, magnetotransport measurements in this low density regime demonstrate high quality transport with well defined integer quantum Hall effects through both vanishing minima of the longitudinal resistance and quantized Hall plateaus, as shown in Fig.\,\ref{algaassb_cap}\,(c). 
High carrier mobility of $\mu \sim 1.0 \times 10^6\,$cm$^{2}$/Vs is demonstrated. 

These results clearly indicate that the gate does not fully deplete the 2DEG in the quantum well. 
Furthermore, the magnetotransport data of Fig.\,\ref{algaassb_cap}(c) does not appear to be consistent with trivial InAs edge conduction, as reported in Ref.\,\onlinecite{mueller:2017}.  
The high mobility maintained at low density suggests the 2DEG is well screened from charged scattering sites that invariably must exist at the dielectric/semiconductor interface.  
Taken as a whole, these data suggest the formation of a mobile screening layer present between the electrostatic gate and the 2DEG.
It is worth noting that similar results were obtained for Sample F where the 2DEG density in the saturation regime was $n\sim 1.3 \times 10^{11}\,$cm$^{-2}$ and equilibrium density of $n\sim 2.5 \times 10^{11}\,$cm$^{-2}$ was observed.

%%%%%%%%%%%%%%%%%%%%%%%%%%%%%%%%%%%%%%%%%%%%%%%%%
\begin{table}[b]
{
  \begin{tabular}{ | c | c | c | c |}
    \hline
    \,\, Sample \,\, &\,Cap Layer \,& Top Barrier & As Cooled $n$  \\ \hline
    \,\,E\,\, &  \,AlGaSbAs ($3\,$nm)\, & $20\,$nm & 9.7 \\  \hline
    \,\,F\,\, & GaSb ($2\,$nm) & $20\,$nm & 9.6 \\  \hline
    \,\,G\,\, & InAs ($2\,$nm)& $20\,$nm & 7.9 \\  \hline
    \,\,H\,\, & AlGaSbAs ($3\,$nm)& $60\,$nm & 3.9 \\ 
    \hline
  \end{tabular}}
  \caption{Samples contain a lower barrier of $20\,$nm of Al$_{0.8}$Ga$_{0.2}$Sb and a $24\,$nm InAs quantum well.
  Sample differences are the thickness of the top barrier (Al$_{0.8}$Ga$_{0.2}$Sb) and the cap layer material.
  The last column is the quantum well density upon initial cooling of the samples with no applied gate bias, in units of $10^{11}\,$cm$^{-2}$.
  }
\label{n_table}
\end{table}

%%%%%%%%%%%%%%%%%%%%%%%%%%%%%%%%%%%%%%%%%%%%%%%%%
 
The gate response of Sample G, grown with a $2\,$nm InAs cap layer, is investigated next. 
Figure \,\ref{inas_cap}\,(a) displays the resistance per square, $r$ (left axis), and conductance, $\sigma$ (right axis), versus $\vg$ obtained by biasing the source and drain contacts of a straight Hall bar of width $150 \,\mu$m and length $1200 \,\mu$m.
From the $n$ versus $\vg$ dependence, we estimate that complete depletion of the 2DEG should occur at $\vg \sim -0.525\,$V, or, possibly, at slightly larger negative voltage if the system were to undergo a metal-insulator-transition \citep{shabani:2014_MIT}.
In the $\vg$ regime beyond pinch off we observe $r>50\,$M$\Omega$/sqr., corresponding to a variation of nearly five orders of magnitude within the measurement range; concurrently $\sigma\rightarrow 0$. 
Deep in the depletion regime accurate determination of resistance is limited by the measurement circuit. 
While we have not explicitly performed an analysis to determine residual edge resistivity, our measurements demonstrate electrostatic control of conductivity in the depleted regime at least comparable to that reported in Ref.\,\onlinecite{mittag:2017} for low density InAs 2DEGs.
For the samples investigated here, \hfo and \alo dielectrics resulted in similar performance. Our results suggest that the use of a thin InAs cap layer coupled with the quaternary Al$_{0.8}$Ga$_{0.2}$Sb$_{0.93}$As$_{0.07}$ buffer facilitate strong electrostatic control of the conductivity.

\subsection{Mobility versus density}
%%%%%%%%%%%%%%%%%%%%%%%%%
\begin{figure}[t]
\vspace{-0.1 in}
\includegraphics[width=0.48\textwidth]{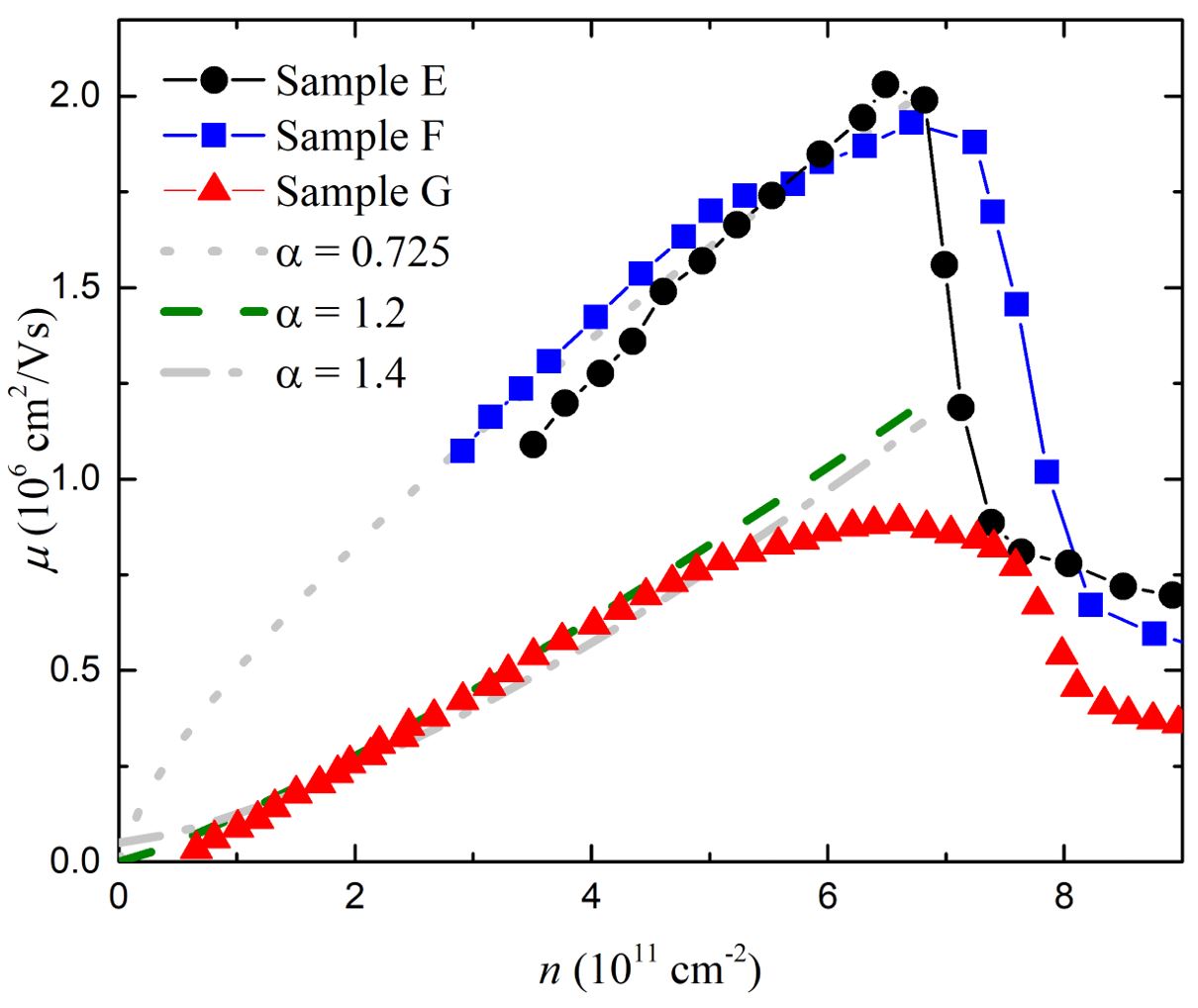}
\vspace{-0.25 in}
\caption{(Color online)
Mobility $\mu$ vs $n$ for different cap layers, Samples E\,-\,G.
Dotted, dashed, and dash-dotted lines correspond to different values of $\alpha$ assuming the relation $\mu\propto n^{\alpha}$.
}
\label{mus}
\end{figure}
%%%%%%%%%%%%%%%%%%%%%%%%%
We continue our investigations by analyzing the impact of the cap layer material on the scattering mechanisms of the 2DEG. 
In particular, we investigate $\mu$ versus $n$ at $T=300\,$mK for Samples E\,-\,G (see Table\,\ref{n_table}), plotted in Fig.\,\ref{mus}.
For all three samples $\mu$ versus $n$ is non-monotonic. 
At low $n$, $\mu$ first increases until $n \sim 6 -7\times 10^{11}$ cm$^{-2}$ and then a sharp decrease is observed for $n\gtrsim 7.5 \times 10^{11}\,$cm$^{-2}$, corresponding to the density at which the second subband of the quantum well starts to be populated.
Samples E and F present nearly identical $\mu$ variation with $n$ with a  maximum mobility of $\mu = 2.1 \times 10^{6}\,$cm$^2$/Vs at $n = 6.7 \times 10^{11}\,$cm$^{-2}$ (magnetotransport at this peak mobility was plotted in Fig.\,\ref{algaassb_cap}\,(a) for Sample E), indicating again that the GaSb and AlGaSbAs cap layers are playing a similar role.
Note that these results appear to compare favorably with Ref.\,\onlinecite{tschirky:2017} in which a peak mobility of $\mu \sim 2.4 \times 10^{6}$ cm$^2$/Vs was measured but at nearly twice the density of $n \sim 12 \times 10^{11}$ cm$^{-2}$ in a $400\,\mu$m Hall bar.
The $\mu$ vs $n$ dependence of Sample G is visibly different in both its dependence and lower peak mobility of $\mu \sim 8.2 \times 10^{5}\,$cm$^2$/Vs, for which the magnetotransport data are presented in Fig.\,\ref{inas_cap}(b).

At low temperature and high carrier density where semiclassical Boltzmann transport theory is valid, the relation $\mu \propto n^{\alpha}$ is often used to characterize the dominate scattering mechanisms \citep{sarma:2013}. 
The value of $\alpha$ reflects the distribution of charged point impurities, which can be localized in a plane parallel to the 2DEG (e.g. delta doping or a nearby surface) or distributed uniformly in bulk of the structure. 
More specifically, $\alpha\rightarrow 0.5$ in the case where the mobility is limited by uniformly distributed 3D ionized background impurities and $\alpha\rightarrow 1.5$ when remote ionized impurity scattering is the limiting factor \citep{sarma:2013}.
Assuming $\mu \propto n^{\alpha}$, for both Samples E and F we estimate $\alpha \sim 0.7-0.75$, similar to Ref.\,\onlinecite{tschirky:2017}; the grey dotted line in Fig.\,\ref{mus} overlaid on the data for $\alpha=0.725$ is a guide to the eye of the estimated relationship.  
At first blush this relationship is surprising considering that the 2DEG resides just $\sim 20\,$ nm below the dielectric/semiconductor interface where a high density of charged defects may be expected.
Repeating the same exercise for Sample G, an exponent $\alpha \sim 1.2-1.4$ is extracted.
In our study the $\alpha$ values have been determined using a log-log scale (not shown here) and $\mu(n)$ has been fitted on a limited $n$ range for $n \leq 6 \times 10^{11}\,$cm$^{-2}$, below the onset of $2^{\mathrm{nd}}$ subband occupation. 
While our values of $\alpha$ are only estimations, it is nevertheless clear that differences in behavior are observed.
Differences in peak mobility value and in $\alpha$ between Samples E and F and Sample G strongly indicate different limiting scattering mechanisms for the structures terminated with Sb-containing layers compared to InAs cap layers. 
These observations suggest that with Sb-based cap layers the 2DEG is well-screened from ionized impurity scattering sites residing at the sample surface that is located only $\sim 20\,$nm away. 
The formation of a weakly conducting layer under negative gate bias between the gate and the 2DEG could explain such behavior. 
A screening layer may also explain the lack of full 2DEG depletion; once enough charge is induced to fully screen the electric field of the gate, the 2DEG density saturates.  
This mechanism is consistent with our observations of a high mobility 2DEG persisting beyond the range of linear gate voltage control.

\subsection{Low density devices}

Sample H incorporates a $60\,$nm top barrier and an Al$_{0.8}$Ga$_{0.2}$Sb$_{0.93}$As$_{0.07}$ cap layer. 
The as-cooled density drops to $n= 3.9 \times 10^{11}$ cm$^{-2}$, more than two times lower than for Sample E where the only difference between the two is an additional $40\,$nm of the same top barrier material. 
As we observed a negligible amount of structural disorder in the buffer layer (see STEM image of Fig.\,\ref{xrd_tem}\,(c)) and our samples lack an intentional doping layer it is reasonable to assume that the charge in the well is primarily supplied by the surface such that an increase in the top barrier thickness results in a decrease in the density transferred to the quantum well, in agreement with our density measurements on Samples E and H.

%%%%%%%%%%%%%%%%%%%%%%%%%
\begin{figure}[t]
\vspace{-0.1 in}
\includegraphics[width=0.48\textwidth]{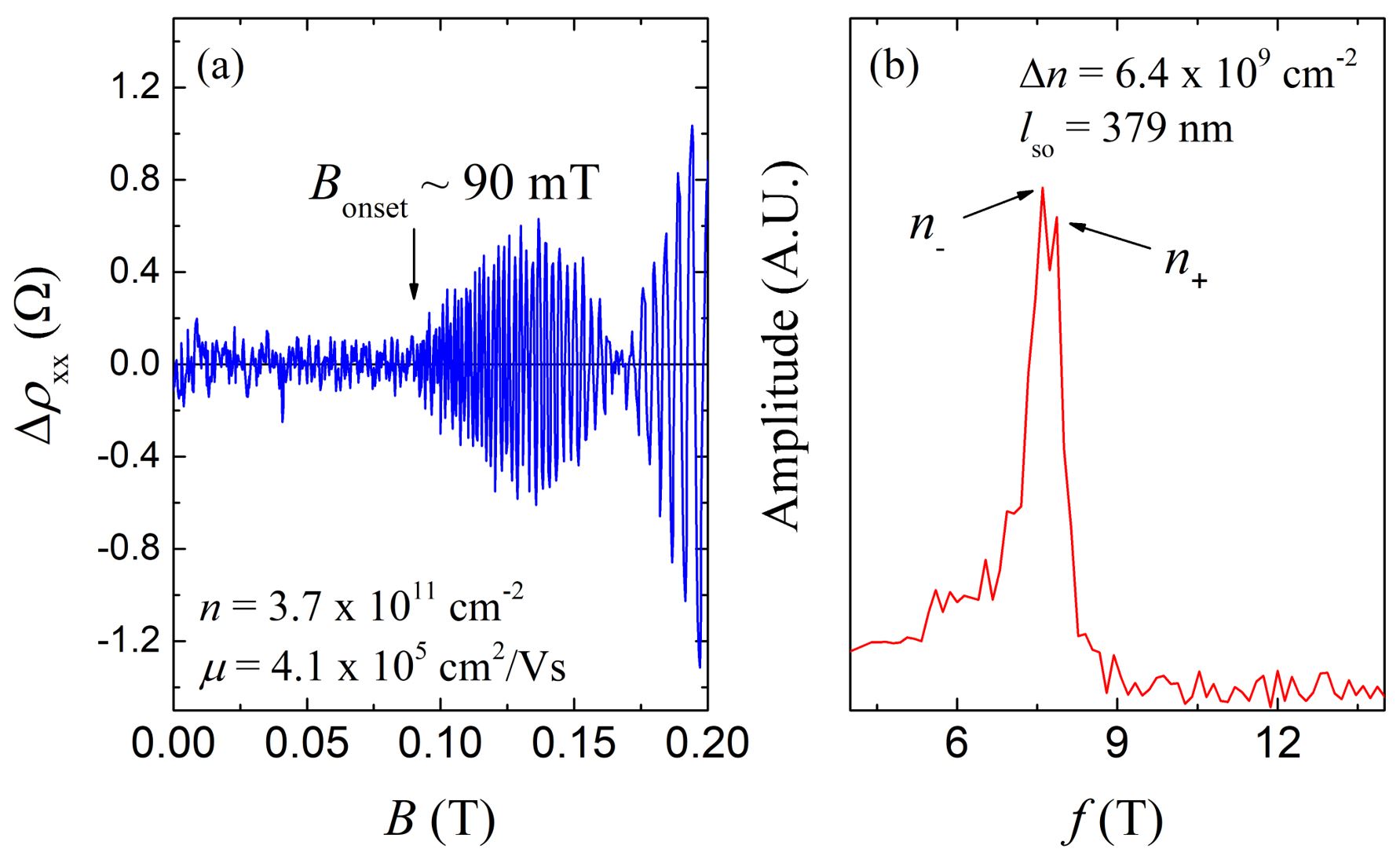}
\vspace{-0.25 in}
\caption{(Color online)
(a) Magnetoresistance at low field for Sample H with removal of a slowly varying background. 
(b) Fast Fourier Transform amplitude of the magnetoresistance of (a) from inverse $B$.
}
\label{fft_fig}
\end{figure}

%%%%%%%%%%%%%%%%%%%%%%%%%
Even at this low starting density, Sample H exhibits a high mobility of $\mu = 4.1 \times 10^{5}\,$cm$^2$/Vs at $n=3.7 \times 10^{11}$ cm$^{-2}$.
In Fig.\,\ref{fft_fig}\,(a) we plot the low field magnetoresistance after removal of a slowly varying background, $\Delta \rho$, where the high quality of the 2DEG is apparent in the onset of the Shubnikov-de Haas oscillations (SdHOs), occurring at $90$ mT.
As a lower bound we estimate the quantum scattering time, $\tq \sim 2\,$ps, from the condition $\oc\tq \sim 1$, where $\oc=\mstar/eB$ is evaluated at the onset of the SdHOs and $\mstar =0.03 \,\mathrm{m}_{\mathrm{e}}$ \citep{shabani:2014_MIT}.
Additionally, $\Delta \rho$ contains a well defined node. 
Fast Fourier Transform of the magnetoresistance of Fig.\,\ref{fft_fig}\,(a) is plotted in Fig.\,\ref{fft_fig}\,(b) as a function of frequency $f$. 
We observe a single split peak indicative of two nearly identical oscillation periods due to zero field spin splitting. The spin up and spin down densities are labeled $n_{+}$ and $n_{-}$, respectively \citep{datta:1990, hatke:2017}, and can be determined from $n_{\pm}=\frac{ef}{h}$.
From these two densities we estimate the Rashba parameter, $\ar=\frac{\Delta n\hbar^{2}}{\mstar}\sqrt{\frac{\pi}{2(\nt-\Delta n)}}=3.39 \times 10^{-12}\,$eV\,m, where $\Delta n = \np-\nn$ and $\nt =\np+\nn$.
Concurrently the spin-orbit length is also estimated through $\lso=\frac{1}{\Delta n}\sqrt{\frac{\nt-\Delta n}{2\pi}}= 379\,$nm.

Interestingly, despite the use of a AlGaSbAs cap layer in Sample H, the $n$ vs $\vg$ variation was completely linear and  we were able to fully deplete the carriers at $\vg=-0.5$V (data not shown here). The as cooled $\vg=0$V density of sample H was only $3.9\times10^{11}$cm$^{-2}$.
Evidently, the reduced negative gate voltage required for depletion of the low density 2DEG in Sample H limits interfacial charge accumulation under the gate.
This behavior contrasts with our previous observations for sample E with higher as-cooled density in which 2DEG density saturated for $\vg \leq -0.7$V without full depletion. Possible explanations for this behavior are discussed below.

\section{Discussion}
For InAs 2DEGs on GaSb substrates, careful consideration of both the buffer layer and cap layer is needed to produce electrical isolation of devices from the substrate and full electrostatic control with lithographic gates while maintaining high 2DEG mobility. 
Only in the case of an InAs cap layer, have we experimentally demonstrated full carrier depletion for near-surface InAs 2DEGs of arbitrary density.

Bandstructure simulations with Nextnano suggest that the conduction band of InAs cap layers may cross the Fermi level.
Keeping the InAs thickness below $2\,$nm is thus important to avoid unintentional population of states at the surface. 
A depletion range on the order of a few hundred mV can be estimated before the Fermi level crosses the valence band of the Al$_{0.8}$Ga$_{0.2}$Sb barriers. 
In simulations, the gate voltage response in structures employing an Al$_{0.8}$Ga$_{0.2}$Sb$_{0.93}$As$_{0.07}$ cap layer appears to be very similar to what is observed with an InAs cap layer and a depletion range of the same order of magnitude is expected. 
The simulated gate voltage response with a GaSb cap behaves differently; the  conduction band minimum of the InAs quantum well and the valence band maximum of the GaSb can coincide with the Fermi level under reverse bias. 
In this case, a hole gas may occur near the semiconductor surface.
Such a hole gas is likely to act as a screening layer between the metallic gate and the 2DEG, limiting gate response for further increase of reverse bias.

While the gate voltage dependence of the bandstructure might explain the incomplete depletion with GaSb cap layer, the situation is less clear for the AlGaSbAs cap layer. 
A major difference between Sb-containing and As-based cap layers is the potential of oxidation and formation of charged states at the surface.  
AlSb-containing alloys are particularly prone to oxidation.
It has been reported that deposition of InAs on GaSb limits the number of defects that were mostly attributed to Sb-oxide at the dielectric/semiconductor interface \citep{Yokoyama:2015}.  
Charges accumulated at the dielectric/semiconductor interface during application of negative gate bias may be expected to impact gate control of 2DEG density.  
Simulations also indicated that the gate voltage value required to reach complete depletion depends on the thickness of the barrier and thus of the as-cooled density. 
The depletion voltage is estimated to be approximately a factor of two smaller for a $60\,$nm barrier ($\sim -0.6$V) compared to the $20\,$nm case ($\sim -1.4$V). 
It is likely that structures with thicker barriers and lower density 2DEGs naturally avoid formation of unintentional charge accumulation layers under reverse bias and are thus easier to control with top gates.

Given the bandstructure evolution with gate voltage and the assumption of charge at the dielectric/semiconductor interface, we have identified two scenarios that could explain the difficulty of depleting high density, near surface 2DEGs having GaSb and AlGaSbAs cap layers. 
GaSb and AlGaSbAs cap layers do appear to limit surface scattering better than InAs capping layers and result in higher mobility.  
Further studies are needed to draw stronger conclusions on the mechanisms responsible.

\section{Conclusion}
In conclusion, we have shown that near-surface InAs 2DEGs grown on GaSb substrates utilizing an Al$_{0.8}$Ga$_{0.2}$Sb$_{0.93}$As$_{0.07}$ buffer and an InAs cap layer exhibit all the properties required to operate quantum devices: proper isolation from the substrate, strong electrostatic control of 2DEG density, and high mobility.  Low density structures, with thicker top barriers, are promising for experiments involving the coupling of superconductivity and integer quantum Hall effect at very low magnetic fields \citep{Lee:2017} or investigations of the fractional quantum Hall effect using readily available superconducting magnets \citep{ma:2017}.

\section{Acknowledgments}
This work was supported by Microsoft Quantum. 
We thank Jan Gukelberger for useful discussions. 

\bibliographystyle{apsrev}
%\bibliography{inas_ah}

\end{document}